\documentclass[twocolumn,prd,showpacs,superscriptaddress,preprintnumbers,nofootinbib]{revtex4-1}
\usepackage{graphicx}
\usepackage{epsfig}
\usepackage{bm}
\usepackage{latexsym,amssymb,amsmath,amsfonts,amssymb,txfonts,pxfonts,wasysym,float}
\usepackage{color}

\newcommand{\beq}[1]{\begin{equation}\label{#1}}
\newcommand{\eeq}{\end{equation}}
\newcommand{\bea}[1]{\begin{eqnarray} \label{#1}}
\newcommand{\eea}{\end{eqnarray}}
\newcommand{\ba}{\begin{array}}
\newcommand{\ea}{\end{array}}

\def\be{\begin{equation}}
\def\ee{\end{equation}}
\def\gs{\mathrel{
   \rlap{\raise 0.511ex \hbox{$>$}}{\lower 0.511ex \hbox{$\sim$}}}}
\def\ls{\mathrel{
   \rlap{\raise 0.511ex \hbox{$<$}}{\lower 0.511ex \hbox{$\sim$}}}}

\newcommand{\postscript}[2]{\setlength{\epsfxsize}{#2\hsize}
   \centerline{\epsfbox{#1}}}

\begin{document}

\title{What IceCube data tell us about neutrino emission from
star-forming galaxies (so far)}
\author{Luis A.~Anchordoqui}
\affiliation{Department of Physics and Astronomy,
Lehman College at CUNY, Bronx NY 10468, USA
}

\affiliation{Department of Physics,
University of Wisconsin-Milwaukee,
 Milwaukee, WI 53201, USA
}

\author{Thomas C. Paul}

\affiliation{Department of Physics,
University of Wisconsin-Milwaukee,
 Milwaukee, WI 53201, USA
}

\affiliation{Department of Physics,
Northeastern University, Boston, MA 02115, USA
}

\author{Luiz~\nolinebreak H.~\nolinebreak M.~\nolinebreak da~\nolinebreak
  Silva}
\affiliation{Department of Physics,
University of Wisconsin-Milwaukee,
 Milwaukee, WI 53201, USA
}

\author{Diego F. Torres}
\affiliation{Institute of Space Sciences (IEEC-CSIC), Campus UAB, Torre C5, 2a planta, 08193 Barcelona, Spain}
\affiliation{Instituci\'o Catalana de Recerca i Estudis Avan\c{c}ats
  (ICREA), Spain}

\author{Brian J. Vlcek}
\affiliation{Department of Physics,
University of Wisconsin-Milwaukee,
 Milwaukee, WI 53201, USA
}

\date{June 2014}
\begin{abstract}
  \noindent Very recently, the IceCube Collaboration reported a flux of
  neutrinos in the energy range $50~{\rm TeV} \alt E_\nu \alt 2~{\rm PeV}$,
  which departs from expectations from atmospheric background at the $5.7\sigma$
  level. This flux is in remarkable agreement with the expected diffuse flux of
  neutrinos from starburst galaxies, and the 3 highest energy events have
  uncertainty contours encompassing some of such systems.  These events, all of
  which have well-measured energies above 1~PeV, exhibit shower topologies, for
  which the angular resolution is about $15^\circ$.  Due to this angular
  uncertainty and the {\em a posteriori} nature of cuts used in our study it is
  not possible to assign a robust statistical significance to this
  association. Using muon tracks, which have angular resolution $< 1^\circ$, we
  compute the number of observations required to make a statistically
  significant statement, and show that in a few years of operation the upgraded
  IceCube detector should be able to confirm or refute this hypothesis. We also
  note that double bang topology rates constitute a possible discriminator among
  various astrophysical sources.
\end{abstract}


\maketitle

In 2012, the IceCube Collaboration famously announced an observation
of two $\sim$~1 PeV neutrinos discovered in a search for the expected
cosmogenic neutrinos~~\cite{Aartsen:2013bka}. The search technique was
refined to extend the neutrino sensitivity to lower
energies~\cite{Schonert:2008is}, resulting in the discovery of an
additional $26$ neutrino candidates with energies between $50$~TeV and
$2$~PeV, constituting a $4.1\sigma$ excess for the combined 28 events
compared to expectations from neutrino and muon backgrounds generated
in Earth's atmosphere~\cite{Aartsen:2013jdh}. Very recently, these
results have been updated~\cite{Aartsen:2014bea}. At the time of
writing, 37 events have been reported in three years of IceCube data
taking (988 days between 2010 -- 2013). The data are consistent with
expectations for equal fluxes of all three neutrino flavors and with
isotropic arrival directions. Moreover, the next to highest energy event has equatorial coordinates
($\alpha = 38.3^\circ, \delta =-67.2^\circ$) and therefore cannot
originate from the Galactic plane.  Assuming a power law spectrum
$\propto E_\nu^{-2}$, the three year data set is consistent with an
astrophysical flux at the level of $3 \times 10^{-8}\, E_\nu^{-2}~{\rm
  GeV}^{-1} \, {\rm cm}^{-2} \, {\rm s}^{-1} \, {\rm sr}^{-1}$, and
rejects a purely atmospheric explanation at
$5.7\sigma$~\cite{Aartsen:2014bea}. Herein we consider the issue of
what the data reported so far may suggest regarding the possibility
that the extraterrestrial neutrinos originate in star-forming
regions~\cite{Anchordoqui:2013dnh}.

Both the neutrino energy spectrum and directional measurements provide
clues about which astrophysical sources may be responsible for
extraterrestrial neutrinos.  We will begin with a discussion of
characteristics of the energy spectrum as it pertains to potential
source candidates, and then move on to the issue of directional
correlations with astrophysical objects. First, however, we should
remind the reader that the three neutrino species $\nu_e$, $\nu_{\mu}$
and $\nu_{\tau}$ induce different characteristic signal morphologies
when they interact in ice producing the Cherenkov light detected by
the IceCube optical modules. The charged current (CC) interaction of $\nu_e$ produces
an electromagnetic shower which ranges out quickly. Such a shower
produces a rather symmetric signal, and hence exhibits a poor angular
resolution of about $15^\circ - 20^\circ$~\cite{Aartsen:2013jdh}.  On
the other hand, a fully or mostly contained shower event allows one to
infer a relatively precise measurement of the $\nu_e$ energy, with a
resolution of $\Delta (\log_{10} E_\nu) \approx
0.26$~\cite{Abbasi:2011ui}.  The situation is reversed for $\nu_{\mu}$
events. In this case, CC interaction in the ice generates a muon which
travels relatively unhindered leaving behind a track.  Tracks
point nearly in the direction of the original $\nu_{\mu}$ and thus
provide good angular resolution of about $0.7 ^\circ$, while the
``electromagnetic equivalent energy'' deposited represents only a
lower bound of the true $\nu_{\mu}$ energy.  The true energy may be up
to a factor 10 larger than the observed
electromagnetic equivalent energy. Finally, $\nu_{\tau}$ CC interactions
may, depending on the neutrino energy, produce ``double bang'' events~\cite{Learned:1994wg},
with one shower produced by the initial $\nu_{\tau}$ collision in the
ice, and the second shower resulting from most subsequent $\tau$
decays.  Separation of the two bangs is feasible for $\nu_{\tau}$
energies above about 3~PeV, while at lower energies the showers tend
to overlap one another~\cite{comment}.

With these points in mind, we now move to the current state of the
neutrino energy measurements. One striking feature of the IceCube
spectrum is that, assuming an unbroken $E_\nu^{-\gamma}, \, \gamma=2$ flux
expected from Fermi acceleration in strong shocks, there is either a
cutoff or a spectral break evident around $2$~PeV.  Notably, there
is no increase in observation rate near $6.3$~PeV, as one would expect
from the Glashow resonance~\cite{Glashow:1960zz}.  This implies that either the acceleration
process dies out at some energy, or that the spectrum is simply
steeper than $\gamma = 2$.  It has been shown elsewhere that an
unbroken power law spectrum with $\gamma = 2.3$ is also reasonably
consistent with the IceCube data~\cite{Anchordoqui:2013qsi}.

In order to ascertain the physical processes which could underlie these spectral
features, let us discuss briefly plausible neutrino production mechanisms. It is
generally thought that extraterrestrial neutrinos are produced via proton
interactions with either photons or gas near the proton acceleration sites,
resulting in pions which in turn generate neutrinos as decay products.  For the
case of neutrino production via $p \gamma$ interactions, the center-of-momentum
energy of the interaction must be sufficient to excite a $\Delta^+$ resonance,
the $\Delta^+ (1232)$ having the largest cross-section. The threshold proton energy
for neutrino production on a thermal photon background of average energy $E_\gamma$ is
\begin{equation} \label{eqn:thresh}
E_{\rm th} = m_\pi ( m_p + m_\pi/2)/E_\gamma \,,
\end{equation}
where $m_\pi$ and $m_p$ are the masses of the pion and proton,
respectively.  Since the proton energy must be about 16 times higher
than the daughter neutrino energies, Eq.~(\ref{eqn:thresh}) implies
photons with energies in the range $\sim 6$~eV should be abundant in
the region of proton acceleration in order to generate $\sim$~PeV
neutrinos. Gamma-ray bursts (GRBs) may be the only astrophysical
objects capable of generating a photon background of the  required energy for
this scenario~\cite{Waxman:1997ti}. Furthermore, production of
neutrinos in the 100~TeV range requires photon energies about an order of
magnitude higher.
In contrast, if neutrinos are produced via
interaction in gas near the acceleration site, the energy threshold
requirement is lifted, as $pp$ interactions generate pions over a
broad range of energies.

\begin{figure}[tbp]
\postscript{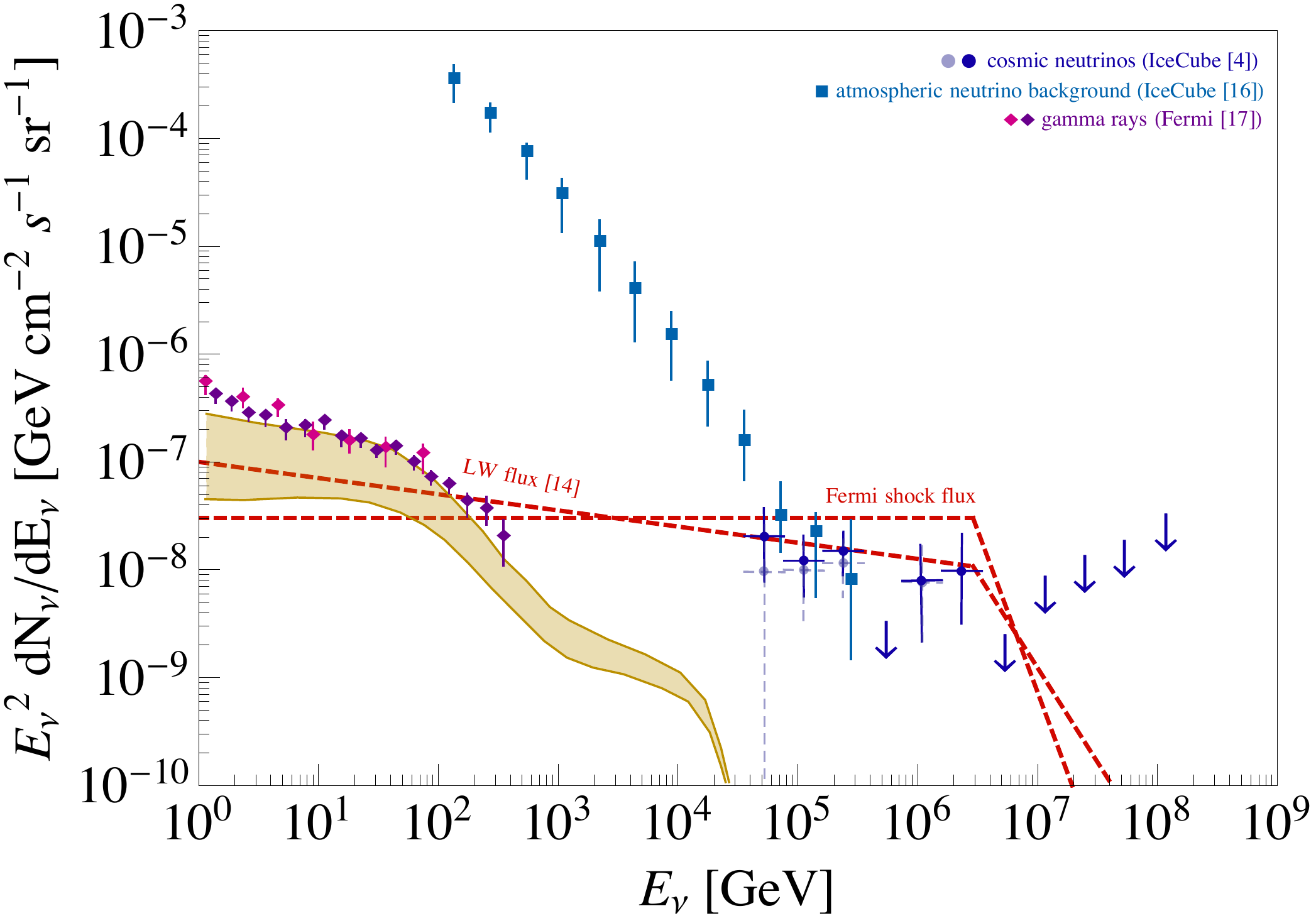}{0.99}
\caption{Neutrino and gamma ray spectra compared to two neutrino
  spectral indices.  The squares show the background from
  atmospheric $\nu_\mu$ events as observed by
  IceCube40~\cite{Abbasi:2010ie}. The circles and arrows show
  the recently reported IceCube flux (points with solid error bars do
  not include prompt background while those with dash error bars do)~\cite{Aartsen:2014bea}.  The
  The  diamonds  are gamma ray flux measurements from Fermi~\cite{Abdo:2010nz}.  The two
  dashed lines correspond to $E_\nu^2 {\rm d}N_\nu / {\rm d}E_\nu =  10^{-7} E^{-0.15}~{\rm GeV} \, {\rm cm}^{-2} \, {\rm s}^{-1}
  \, {\rm sr}^{-1}$ and $E_\nu^2 {\rm d}N_\nu / {\rm d}E_\nu  = 3
  \times 10^{-8}~{\rm GeV} \, {\rm cm}^{-2} \, {\rm s}^{-1} \, {\rm
    sr}^{-1}$, with the spectrum steepening above about 2~PeV to
  $\gamma = 3.75$ and $\gamma = 5.0$, respectively.  For these two
  neutrino fluxes, the associated predictions for the gamma ray fluxes
  after propagation are displayed as the upper and lower bounds of the
  shaded region~\cite{Murase:2013rfa} . Note that the spectral index
  $\gamma = 2.15$ at injection agrees well with both the Fermi-LAT and
  IceCube measurements.}
\label{fig:uno}
\end{figure}

Extending previous multifrequency studies of individual galaxies~\cite{Pavlidou:2001gi},
Loeb and Waxman (LW)~\cite{Loeb:2006tw} showed in 2006 that
starburst galaxies constitute a compelling source for efficient neutrino
production up to $\sim 0.3$~PeV, and possibly beyond, though for
energies exceeding 1~PeV the predictions are quite uncertain.  For energies up to
$\sim 1$~PeV, the LW analysis predicts a spectral index $\gamma = 2.15 \pm 0.10$
which accurately fits the IceCube data, and indeed predicts an observation rate for
$E_\nu$ of $10^{1.5 \pm 0.05}$ for a $1~{\rm km^3}$ detector, in line with
the rate subsequently observed by IceCube. Neutrino production from $\pi^\pm$ decays must be
accompanied by a corresponding flux of gamma rays from decays of $\pi^0$'s
produced in the $pp$ interactions, providing a robust cross-check of the pion
production rate and corresponding neutrino spectrum.  A spectrum steeper than
$\gamma \sim 2.2$ leads to an overproduction of gamma rays compared to
measurements by Fermi-LAT~\cite{Murase:2013rfa}, indicating that a soft unbroken $\gamma
= 2.3$ spectrum is implausible for extragalactic sources. Thus, it seems that a
cutoff or suppression must be at play. All in all, the starburst source hypothesis
together with a steepening of the spectrum to at least $\gamma = 3.75$ above 3~PeV
fits well to the IceCube data and satisfies the constraints from gamma ray
observations, as shown in Fig.~\ref{fig:uno}.

We now discuss how double bang topologies may serve as a discriminator
among possible astrophysical sources powered by highly relativistic winds. Extraterrestrial neutrino
production proceeds via the decay chain
\begin{eqnarray}\label{eqn:chain}
\pi^+ \to & \mu^+ & \nu_\mu ~~~~~~~~~~~~~~~~{\rm (and \, the \,
  conjugate \, process)} \,. \\
                & \reflectbox{\rotatebox[origin=c]{180}{$\Rsh$}} &
                 e^+ \, \bar \nu_\mu  \, \nu_e \nonumber
\end{eqnarray}
This decay chain may be complete in the sense that both decays indicated in
Eq.~(\ref{eqn:chain}) occur without significant change in the $\mu$ energy, or
it may be incomplete, in which case the $\mu$ suffers possibly catastrophic
energy loss before decay. For the case of a complete decay chain, each neutrino
carries on average about $1/4$ of the parent pion energy. If the $\mu$ radiates
away energy before it decays, the $\nu_{\mu}$ from the first decay will still
carry on average $1/4$ of the $\pi^{\pm}$ energy, while the other two neutrinos
will emerge with less than the nominal $1/4$ of the parent pion energy. In such
a scenario it is conceivable that the first $\nu_\mu$ in the chain can be
produced above 3 PeV, whereas $\bar{\nu}_e$ may not reach beyond $2~{\rm PeV}$,
and in particular may not be able to reach the energy required to interact at
the Glashow resonance.

We now discuss the muon energy loss quantitatively by exploiting the
observation of gamma rays accompanying the neutrino flux.  In the case
of muons with energies in excess of $~1$~PeV, energy losses are
dominated by synchrotron radiation. The synchrotron loss time is
determined by the energy density of the magnetic field in the wind
rest frame.  Defining $\tau_{\mu,{\rm syn}}$ as the characteristic muon cooling
time via synchrotron radiation and $\tau_{\mu,{\rm decay}}$ as the muon
decay time, it is necessary that $\tau_{\mu,{\rm syn}} < \tau_{\mu,{\rm decay}}$
in order for the decay chain to be complete.  $\tau_{\mu,{\rm syn}} \sim
\tau_{\mu,{\rm decay}}$ determines a critical energy $E_\mu^{\rm syn}$ at which
energy losses begin to affect the decay chain.  For the characteristic parameters of a GRB wind, the
maximum energy at which all neutrinos in the decay chain have on
average 1/4 of the pion energy is
\begin{equation}
E_\nu^{\rm syn} \approx \frac{1}{3} \, E_\mu^{\rm syn}  \sim \frac{1}{3}  \frac{\Gamma_{2.5}^4 \, \Delta
t_{-3}}{L_{52}^{1/2}}~{\rm PeV},
\label{muonloss}
\end{equation}
where $\Gamma = 10^{2.5} \Gamma_{2.5}$ is the wind Lorentz factor, $L
= 10^{52} L_{52}~{\rm erg/s}$ is the kinetic energy luminosity of the
wind, and $\Delta t = 10^{−3} \Delta t_{-3}$ s is the observed
variability time scale of the gamma-ray signal~\cite{Rachen:1998fd}.
Equation~\ref{muonloss} is also valid for neutrinos produced in
blazars. In this case, $\Delta t \sim 10^4~{\rm s}$, $\Gamma \sim 10$,
and $L \sim 10^{47}$~erg/s, yielding $E_\nu^{\rm syn} \sim 1~{\rm
  EeV}$.  For starburts, the galactic wind is non-relativistic and the
magnetic field is small enough to render synchrotron losses negligible
in comparison. In summary, for GRBs, the muon cooling is sufficient to
influence the decay chain in such a way as to affect the flavor ratios
at PeV energies, whereas for blazar and starbursts the decay chain is
only affected for muon energies $\gg 10$~PeV.  Note
that for GRBs, $\Delta t_{-3} \sim 1$ constitutes a lower bound, and hence
the consequences discussed herein may require some fine-tuning of the
parameters of Eq.~(\ref{muonloss}).

It is  nonetheless worth noting some potential consequences of the above
hypotheses. As noted elsewhere $p \gamma$ interactions produce fewer $\bar
\nu_e$ than $pp$ ineractions~\cite{Anchordoqui:2004eb}. Indeed, most of the
$\bar \nu_e$ flux originates via oscillations of $\bar \nu_\mu$ produced via
$\mu^+$ decay. For production of $E_\nu \agt 1~{\rm PeV}$ in GRBs, the $\nu_\mu$
in the chain of Eq.~(\ref{eqn:chain}) is more energetic than the $\bar
\nu_\mu$. This may suggest that the softening of the spectral index takes place
at different energies for neutrino and antineutrino fluxes. If this were the
case, at production the high energy end of the GRB flux would be dominated by
$\nu_\mu$ produced via $\pi^+$ decay. As described previously, however, IceCube
can measure only lower bounds for the muon energies. As it turns out, IceCube
has recently recorded a $\nu_{\mu}$ with a minimum energy of 0.5
PeV~\cite{Karle}, but which may have an energy as much as 10 times higher.  If
this is indeed the case, it could indicate a high energy muon from the first
decay of Eq.~(\ref{eqn:chain}).  We can also speculate on more potentially
convincing observations which may emerge in the future.  Assuming maximal
$\nu_\mu - \nu_\tau$ mixing, observation of a high energy $\nu_{\mu}$
 may imply eventual observation of a high energy $\nu_{\tau}$,
which above about 3~PeV would exhibit the distinctive double bang topology
discussed above.  Note that some fine tuning of the model presented
  here may be required for such events to manifest. In particular, the
  $\mu^{\pm}$ cooling time of Eq.~(\ref{muonloss}) must  be smaller than the
  $\mu^{\pm}$ decay time in order to prevent the $\bar \nu_e$ from reaching the
  Glashow resonance (thus far not observed).  Further, the $\pi^{\pm}$ cooling
  time must exceed its lifetime in order to produce a $\nu_\tau$ above $\sim
  3$~PeV. Further, as this is a phenomenological exercise, we have neglected
  possible experimental effects. As such, this study is not meant to make a
  concrete prediction, but rather to point out that {\em if} such double bang
  topologies are observed in the future while  the Glashow resonance is not, it would
  provide a valuable piece to the puzzle of extraterrestrial neutrino origins,
  favoring the GRB hypothesis over  the blazar or starburst ones, each
  of which would require implausible fine-tuning to be consistent with
  observation.

\begin{figure}[tbp]
\postscript{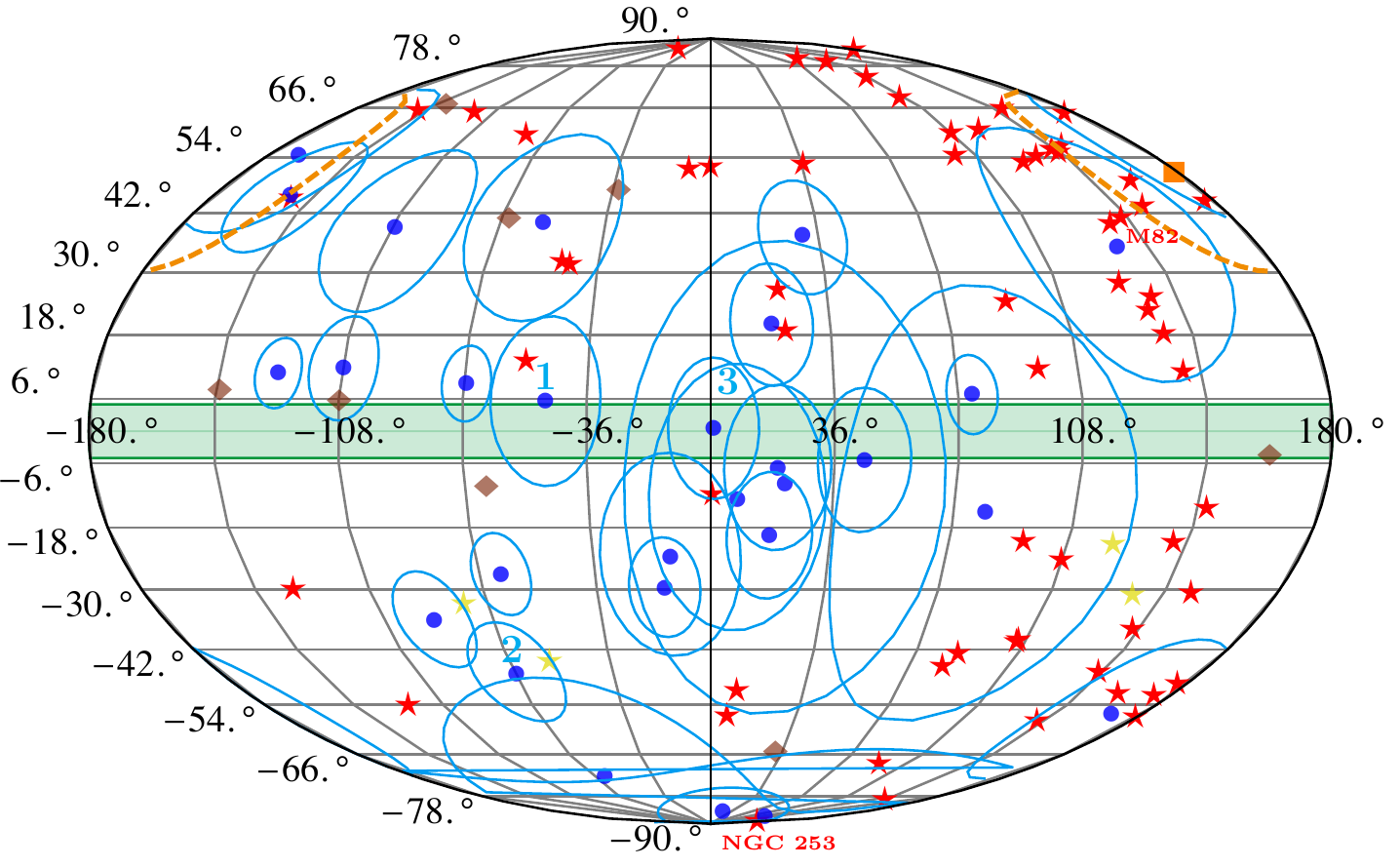}{0.99}
\caption{Comparison of IceCube event locations~\cite{Aartsen:2014bea}
  with star-forming galaxies~\cite{Ackermann:2012vca} and the
  ultrahigh energy cosmic ray hot-spot reported by the TA
  Collaboration~\cite{Abbasi:2014lda} in a Mollweide
  projection. The 27 shower events (circles) and 8 track events
  (diamonds) reported by the IceCube Collaboration. The 3 highest
  energy events are labeled 1, 2, 3, from high to low,
  respectively. The red stars indicate the 64 star-forming galaxies
  beyond the Local Group. The 4 yellow stars indicate
  galaxies in the local group. NGC 253 and M82, our two closest
  starbursts, are labeled. The shaded band delimits the Galactic
  plane. The square in the upper right marks the center of the TA
  hot-spot, with the surrounding dashed line indicating its $20^\circ$
  extent.}
\label{fig:dos}
\end{figure}

Now, since starburst galaxies are plausible source candidates,
consistent with the neutrino energetics observed so far, the next
obvious step is to check whether there are any correlations with the
positions of starburst galaxies and the observed neutrino arrival
directions. Before proceeding we note that hypernovae, which may well
be responsible for sub-PeV to PeV neutrino
emissions~\cite{He:2013cqa}, are present in starburst galaxies as well
as other star forming regions, though the rate of occurrence is higher
in starburst galaxies.  To test the hypothesis that star forming
regions correlate with the IceCube events, we have employed the list
of star-forming regions compiled by the Fermi-LAT Collaboration
~\cite{Ackermann:2012vca}, which includes 64 of the 65 sources of the
HCN survey~\cite{Gao:2003qm} as well as the local galaxies (SMC, LMC,
M31, and M33).  The HCN survey is, to date, the most complete study of
galaxies with dense molecular gas content. It includes nearly all the
IR-bright galaxies in the northern sky $(\delta \geq -35^\circ$) with
strong CO emission, as well as additional galaxies taken from other
surveys. Objects within the Galactic latitudes $|b| < 10^\circ$ are
not included in the survey due to diffuse emission from the Galactic
plane.

A comparison among all of the IceCube events and the star-forming
galaxy survey is shown in Fig.~\ref{fig:dos}. Not surprisingly given
the size of the localization error, there are a few coincidences,
among them the two nearby starbursts M82 and NGC 253 (observed in
gamma-rays~\cite{Acero:2009nb,DomingoSantamaria:2005qk} which are
considered to be possible ultrahigh energy cosmic ray
emitters~\cite{Anchordoqui:1999cu}).  The highest energy event
correlates with NGC 4945, the second highest with the SMC, and the
third highest correlates with IRAS 18293-3413. However, none of the
track topologies correlates with an object in the survey.

To estimate the number of $\nu_\mu$ required to make a statistically
significant statement, we have run $10^6$ simulations with 68 sources
and computed the fraction correlating by chance with $1^\circ$
circular regions of the sky.  Of these, 90\% of the simulations show 0
correlations.  If future observations contain 5 or more $\nu_\mu$
events which correlate with the 68 sources in the survey, an
association by chance will be excluded at more than 99\%
CL~\cite{Feldman:1997qc}.

For $\nu_\mu$ events, the equivalent electromagnetic energy represents
only a lower bound on the true neutrino energy. Consequently, escaping
the background region requires setting a cut on the electromagnetic
equivalent energy $\simeq 0.5~{\rm PeV}$. This threshold is arrived at
via the following argument. Figure 1 shows that at $E_\nu = 1~{\rm
  PeV}$ the background from prompt emission is negligible. Since the
muon neutrino energy is at least 2 times the inferred electromagnetic
equivalent energy, the proposed cut produces a virtually
background-free sample.  Since 1 such event has already been recorded,
we might guess an observation rate of 1 event every $\sim$~2 years,
indicating a long wait with the current $1~{\rm km}^3$
configuration. Next generation IceCube, which could increase the
instrumented volume by up to an order of magnitude (but with larger
string spacing), will therefore be greatly beneficial for this study,
as well as other correlation analyses.

We conclude with one additional observation. It was recently
noted~\cite{Fang:2014uja} that the ultrahigh energy cosmic ray  hot-spot
reported by the TA Collaboration~\cite{Abbasi:2014lda} correlates with
2 of the 28 events initially reported by the IceCube
Collaboration~\cite{Aartsen:2013jdh}, with a statistical significance
of around $2\sigma$.  In the newer IceCube data (the 37 event sample~\cite{Aartsen:2014bea})
there is one additional shower event which
correlates with the TA hot-spot, as shown in Fig~\ref{fig:dos}. The
hot-spot also contains an abundance of star-forming regions and is near M82.\\

This work was supported in part by the US NSF grants CAREER PHY-1053663
(LAA) and PHY-1205854 (TCP), the NASA grant NNX13AH52G (LAA, TCP), the UWM Physics 2014 Summer Research Award
(LHMdaS),  the grant AYA2012-39303 (DFT), and the
UWM RGI (BJV).


\begin{thebibliography}{99}
\bibitem{Gao:2003qm}
  Y.~Gao and P.~M.~Solomon,
  Astrophys.\ J.\  {\bf 606}, 271 (2004)
  [astro-ph/0310339].
The only galaxy from the HCN survey not included in the sample is
  NGC 6921.
\end{thebibliography}
\end{document}